\documentclass[11pt,a4paper]{article}
\usepackage{color,graphics,amsmath,epsfig,rotating,cite}
\usepackage{latexsym}
\usepackage{amssymb}
\usepackage{bbm}
\usepackage{graphicx}
\usepackage{subfig}
\usepackage{wrapfig}
\usepackage{psfrag} 
\usepackage{xcolor}
\usepackage{eurosym}
\usepackage{feynmp}
\begin{document}

\setlength{\unitlength}{1mm}
\renewcommand{\arraystretch}{1.4}


\def\micromegas   {{\tt micrOMEGAs}}
\def\suspect   {{\tt Suspect}}
\def\micro{{\tt micrOMEGAs}}
\def\calchep   {{\tt CalcHEP}}
\def\lanhep   {{\tt LanHEP}}

\def\hinv{{\rm BR}(h\rightarrow {\rm inv})}

\def\ma{M_A}
\def\ra{\rightarrow}
\def\snl{\tilde{\nu}_L}
\def\mneut{M_{\tilde{\chi}^0_1}}
\def\mchi{M_{\tilde{\chi}^0_i}}
\def\mneutt{M_{\tilde{\chi}^0_2}}
\def\mneuth{M_{\tilde{\chi}^0_3}}
\def\mneutf{M_{\tilde{\chi}^0_4}}
\def\mchar{M_{\tilde{\chi}^+_1}}
\def\mchart{M_{\tilde{\chi}^+_2}}
\def\msel{M_{\tilde{e}_L}}
\def\mser{M_{\tilde{e}_R}}
\def\mslo{M_{\tilde{\tau}_1}}
\def\mslt{M_{\tilde{\tau}_2}}
\def\msul{M_{\tilde{u}_L}}
\def\msur{M_{\tilde{u}_R}}
\def\msdl{M_{\tilde{d}_L}}
\def\msdr{M_{\tilde{d}_R}}
\def\msto{M_{\tilde{t}_1}}
\def\mstt{M_{\tilde{t}_2}}
\def\msbo{M_{\tilde{b}_1}}
\def\msbt{M_{\tilde{b}_2}}
\def\sw{s_W}
\def\cw{c_W}
\def\ca{\cos\alpha}
\def\cb{c_\beta}
\def\sa{\sin\alpha}
\def\sb{s_\beta}
\def\tb{\tan\beta}
\def\ssi{\sigma^{SI}_{\lsp N}}
\def\si{\sigma^{SI}}
\def\sip{\sigma^{SI}_{\chi p}}
\def\ssd{\sigma^{SD}_{\chi N}}
\def\sd{\sigma^{SD}}
\def\sdp{\sigma^{SD}_{\chi p}}
\def\sdn{\sigma^{SD}_{\chi n}}
\def\msl{M_{\tilde l}}
\def\msq{M_{\tilde q}}
\def\bsg{B(b\rightarrow s\gamma)}
\def\bsmu{BR(B_s\rightarrow\mu^+\mu^-)}
\def\btau{R(B\rightarrow\tau\nu)}
\def\Omg{\Omega h^2}
\def\sip{\sigma^{SI}_{\chi p}}
\def\amu{\delta a_\mu}
\def\neuto{\tilde\chi^0_1}
\def\neuti{\tilde\chi^0_i}
\def\neutt{\tilde\chi^0_2}
\def\neuth{\tilde\chi^0_3}
\def\neutf{\tilde\chi^0_4}
\def\chargi{\tilde\chi^+_i}
\def\charg{\tilde\chi^+_1}
\def\chargt{\tilde\chi^+_2}
\def\gluino{\tilde{g}}
\def\ul{\tilde{u}_L}
\def\ur{\tilde{u}_R}
\def\stau{\tilde{\tau}}
\def\sl{\tilde{l}}
\def\sq{\tilde{q}}
\def\bone{B^1}
\def\sneutrino{\tilde\nu}
\def\msnu{M_{\tilde\nu_R}}
\def\anu{A_{\tilde\nu}}
\def\sn{\sin\theta}
\def\mzp{M_{Z'}}
\def\azz{\alpha_{Z}}
\def\tesix{\theta_{E_6}}
\def\beq{\begin{equation}}
\def\eeq{\end{equation}}
\def\gcinq{\gamma_5}
\def\wino{\tilde{W}}
\def\bino{\tilde{B}}
\def\binop{\tilde{B'}}
\def\cw{c_W}
\def\sw{s_W}
\def\gggg{R_{gg\gamma\gamma}}
\def\ggtau{R_{gg\tau\tau}}
\def\bbtau{R_{bb\tau\tau}}

\newcommand{\ablabels}[3]{
 \begin{picture}(100,0)\setlength{\unitlength}{1mm}
  \put(#1,#3){\bf (a)}
  \put(#2,#3){\bf (b)}
 \end{picture}\\[-8mm]
}

\newcommand{\gsim}{\;\raisebox{-0.9ex}      {$\textstyle\stackrel{\textstyle >}{\sim}$}\;}


\begin{flushright}
  \vspace*{-18mm}
  Date: \today
\end{flushright}
\vspace*{2mm}

\begin{center}

{\Large\bf The Higgs boson in the MSSM in light of the LHC} \\[8mm]

{\large  D.~Albornoz V\'asquez$^1$, G.~B\'elanger$^2$, R.~M.~Godbole$^3$ and A. Pukhov$^4$}\\[4mm]
{\it 1) CNRS, UMR7095, Institut d'Astrophysique de Paris, F-75014 Paris, France\\
2) LAPTH, Univ. de Savoie, CNRS, B.P.110, F-74941 Annecy-le-Vieux Cedex, France\\
3) Centre for High Energy Physics, Indian Institute of Science, Bangalore, 560 012, India\\
4) Skobeltsyn Inst. of Nuclear Physics, Moscow State Univ., Moscow 119992, Russia
 }\\[4mm]

\end{center}

\begin{abstract}
We investigate the expectations for the light Higgs signal in the MSSM in different search channels at the LHC. 
After taking into account dark matter and flavor constraints in the MSSM with eleven free parameters, we show that the light Higgs signal in the $\gamma\gamma$ channel is expected to be at most at the level of the SM Higgs, while the $h\ra b\bar{b}$ from W fusion and/or the $h \ra\tau\bar\tau$ can be enhanced. 
For the main discovery mode, we show that a strong suppression of the signal occurs in two different cases: low $M_A$ or large invisible width. A more modest suppression is associated with the effect of light supersymmetric particles. Looking for such modification of the Higgs properties and searching for supersymmetric partners and pseudoscalar Higgs offer two complementary probes of supersymmetry. 
\end{abstract}

\section{Introduction}

The search for the Higgs boson is the primary goal of the Large Hadron Collider (LHC). The standard model (SM) Higgs  in the mass range $141~{\rm GeV}<M_h<475~{\rm GeV}$ has been excluded by combined CMS and ATLAS analyses
with integrated luminosities from $1$ to $2.3 fb^{-1}$~\cite{CMS_ATLAS_Higgs}. In this mass range the primary search mode is $H\ra WW^{(*)}\ra l\nu l\nu$.
For lower Higgs masses the main search channel  is $gg\ra H\ra \gamma\gamma$ where both the production and the decay are loop-induced processes. Current limits range from roughly 1-4 times the SM cross section~\cite{CMS_ATLAS_Higgs,CMS_higgs21}.\footnote{The ATLAS collaboration has very  recently reported an excess in this channel that could be compatible with a Higgs with a mass of 126GeV~\cite{Higgs125} while  the CMS collaboration was not able to rule out the region between $117~{\rm GeV}<M_h<127~{\rm GeV}$ because of an excess of events. }
The channels  $H\rightarrow WW^{(*)}\ra l\nu l\nu$ as well as $H\rightarrow ZZ^{(*)}\ra llll$ still play a role in setting the global exclusion limit especially for masses near 140 GeV.
Other search channels include associated Higgs and vector boson production followed by $H\ra b\bar{b}$ with a limit on the cross section relative to the SM one, $\sigma/\sigma_{SM}< 6$ for a Higgs boson mass in the range from 110 to 130 GeV as well as $H\ra\tau\tau$ from vector boson and gluon fusion 
with a limit on $\sigma/\sigma_{SM}\approx 4-6$ for the Higgs in the range from 110 to 130 GeV.

These results however do not yet constrain the light Higgs in the Minimal Supersymmetric Standard Model (MSSM). Indeed in the MSSM the radiative corrections can only raise the lightest Higgs mass to about 135~GeV. As for the SM Higgs, searches for the light Higgs in the MSSM rely mainly on $gg\ra h\ra \gamma\gamma$ with some contribution from the $h\rightarrow WW$ channel in the upper region of the mass range. 
The combined ATLAS and CMS limits for the $h\gamma\gamma$ channel exclude $\sigma/\sigma_{SM}<3.1$ in the mass range $110-121$~GeV and $\sigma/\sigma_{SM}<1.8$ in the $121-131$~GeV range where $\sigma(\sigma_{SM})$ is the production cross section times branching ratio in the MSSM (SM).
For the higher mass range, the collaboration have indeed become sensitive to $\sigma/\sigma_{SM}\approx 1$ 
with the integrated luminosity ${\cal L}=5fb^{-1}$~\cite{Higgs125} as was projected in~\cite{CMS_higgs21}. The channel $H/A/h\rightarrow \tau\tau$ where the Higgses are produced via gluon fusion, or from b-quarks is the preferred channel for the search for the heavy Higgs doublet. This channel has in fact been used to set powerful limits on the heavy Higgs of the MSSM which has enhanced couplings at large values of $\tan\beta$ ~\cite{CMS_higgs20}. Although more modest, the light Higgs couplings can also be enhanced. 
In the mass range relevant for the light Higgs, limits on $\sigma/\sigma_{SM}$ range from 6-10~\cite{CMS_higgs20}.
Note however that $\sigma_{SM}$ in the denominator includes only WW and gg fusion production processes.

The predictions for the supersymmetric Higgses can differ significantly from the SM ones for both production and decay. 
The Higgs couplings to vector bosons or to fermions (in particular to b-quarks) differ from their SM value in the limit where the mass of the pseudoscalar Higgs, $M_A$ is low~\cite{Djouadi:2005gj, Drees:book,Baer:2006rs}. Furthermore the loop-induced couplings $hgg$ and $h\gamma\gamma$ can receive important supersymmetric contributions from the chargino and stop sector.
Finally the mass of the lightest neutralino could be below $M_h/2$ thus leading to invisible decays of the light Higgs and to reduced rates for all visible Higgs decays. Note that for kinematical reasons this occurs mainly in models with non-universal couplings where very light neutralinos are allowed (in the constrained MSSM, LEP limits imply a neutralino above 46~GeV~\cite{Nakamura:2010zzi}).
 
 The implications of supersymmetry on Higgs searches have been extensively studied over the years. 
 The effect of top squarks on loop-induced process on the Higgs signal 
 showed that large corrections can be found especially for light stops ~\cite{Djouadi:1998az,Belanger:1999pv,Carena:2002qg,Kinnunen:2005aq,Dermisek:2007fi,Low:2009nj}. The effect of charginos on the two photon width of the Higgs was shown to be more modest ~\cite{Djouadi:1996pb,Belanger:2000tg}.
Changes in the $h\ra b\bar{b}$ partial width due to supersymmetric corrections can enhance or suppress the branching fraction into two-photons~\cite{Carena:2002qg, Djouadi:2005gj}. The modification of the $h\ra b\bar{b}$ partial width by introducing an extra singlet that mixes with the Higgs can also enhance the two-photon branching fraction in the NMSSM~\cite{Ellwanger:2010nf}.
 More generally, in the MSSM with new degrees of freedom (BMSSM), the 
 modification of the Higgs couplings were shown to lead to potentially suppressed or enhanced signals in all channels~\cite{Carena:2010cs,Carena:2011dm,Boudjema:2011xf}.

Invisible Higgs decays were examined in the framework of the MSSM with non-universal gaugino masses~\cite{Belanger:2000tg,Belanger:2001am}
as well as in the R-parity violating MSSM ~\cite{Hirsch:2005wd,Accomando:2006ga}, in the MSSM with a mixed sneutrino~\cite{ArkaniHamed:2000bq,Thomas:2007bu,Belanger:2010cd} in the NMSSM~\cite{Ellwanger:2005uu,Ellwanger:2009dp} or in the NMSSM with sneutrino dark matter~\cite{Cerdeno:2011qv}. 
For the latter the Higgs is made invisible by decaying into modes not searched for in the standard channels, typically into other light Higgses.

In addition to the above,  invisible Higgs decays exist   in many other models with a light dark matter candidate, including those with a scalar dark matter~\cite{Mambrini:2011ik,Ghosh:2011qc,Andreas:2010dz}, a hidden sector ~\cite{Lebedev:2011iq} or in higher dimensional models ~\cite{Giudice:2000av,Battaglia:2004js}. More generally the Higgs can become invisible in models with extra singlets~\cite{vanderBij:2006pg,vanderBij:2006ne} or with an extra neutrino~\cite{Belotsky:2002ym}.
  Dedicated searches for the invisible Higgs were advocated at the LHC in either vector boson fusion~\cite{Eboli:2000ze,Davoudiasl:2004aj} or in associated vector boson production~\cite{Godbole:2003it,Davoudiasl:2004aj}, see also ~\cite{Cavalli:2002vs,Assamagan:2004mu,Buttar:2006zd}. Methods for measuring the invisible width were suggested~\cite{Low:2011kp,Boos:2010pu}.

In this paper we reexamine the case of the Higgs production and decays in the MSSM with eleven free parameters.
We use a MCMC approach to constrain the parameter space and take into account constraints from B-physics, $(g-2)_\mu$, neutralino relic density, LEP limits on supersymmetric particles, on the Z invisible width and on associated neutralino LSP production as well as limits from Higgs searches at colliders. We further impose additional constraints from dark matter searches : direct detection limits from XENON100~\cite{Aprile:2011hi} and the photon flux from dwarf Spheroidal galaxies (dSph) measured by FermiLAT~\cite{Abdo:2010ex}.
We consider only scenarios where the LSP is a neutralino since a sneutrino LSP is severely constrained by direct detection limits~\cite{Falk:1994es}.
We then examine the consequences for $gg\ra h\ra \gamma\gamma$, $h\ra\tau\bar\tau$ and $h\ra b\bar{b}$ decays as well as for invisible Higgs decays. 
We find that in general the $gg\ra h\ra \gamma\gamma$ is comparable or somewhat suppressed as compared to the SM Higgs of the same mass. However large suppressions can also occur, 
we show that the suppressions are associated with the presence of some light particles thus highlighting the complementarity between SUSY and light/heavy Higgs searches. 

This paper is organized as follows. Section 2 presents briefly the model and the constraints used in the fit. The observables linked to the light Higgs are presented in section 3. Our resuts are summarised in section 4 for each main Higgs decay channel. Benchmarks are also proposed. 
Section 5 contains our conclusions.

\section{Model and constraints}

We consider the MSSM with eleven free parameters defined at the electroweak scale, as in~\cite{Vasquez:2011yq}:

\begin{equation}
M_1, \; M_2, \; M_3, \; \mu, \; \tan\beta , \; M_A, \nonumber\\
M_{\tilde{l}_L}, \; M_{\tilde{l}_R}, \; M_{\tilde{q}_{1,\;2}}, \; M_{\tilde{q}_3}, \; A_t. \nonumber
\end{equation}

We assume minimal flavour violation, two common soft masses $M_{\tilde{l}_L}$ and $M_{\tilde{l}_R}$ for left-handed and right-handed sleptons, equality of the soft squark masses between the first and second generations, $M_{\tilde{q}_{1,\;2}}$, while the mass of the third generation squarks is kept as an independent free parameter $M_{\tilde{q}_3}$. We allow for only one non-zero trilinear coupling, $A_t$. The gaugino masses $M_1, M_2$ and $M_3$ are free parameters as well. In particular this allows to have $M_1\ll M_2$, implying a light neutralino much below the EW scale. The ratio of the doublet Higgs VEV's $\tan\beta$, the Higgs bilinear term, $\mu$, and the pseudoscalar mass $M_A$ are the remaining free parameters. 
This reduced set of parameters as compared to the pMSSM with 19 parameters captures the main feature of the light neutralino dark matter (DM) and allows in particular to explore thoroughly the region where the Higgs decays invisibly. However assuming a common third generation squark could have some influence on the Higgs predictions. Indeed the stops tend to be rather heavy because of the constraints from flavor physics on the sbottom sector thus limiting the impact of the stop sector on the loop-induced Higgs decays. For other studies and global fits of the MSSM with 19 free parameters see~\cite{Sekmen:2011cz, Arbey:2011un,Berger:2008cq,AbdusSalam:2009qd}.

\subsection{Scanning method} \label{sec:Meth}

In order to thoroughly scan the parameter space we used a Markov Chain Monte-Carlo (MCMC), first presented in~\cite{Vasquez:2010ru}, the data set is the same one used in ~\cite{directionnelle}. The code consists on a Metropolis-Hastings algorithm, and is based on micrOMEGAs2.4~\cite{Belanger:2006is,Belanger:2008sj,Belanger:2010gh} for the computation of all observables. The supersymmetric spectra are calculated with SuSpect~\cite{Djouadi:2002ze}. 

Each point is generated by making a random step with a normal variation from the previous point in each dimension. Then, we compute its total prior $\mathcal{P}$, total likelihood $L$ and total weight $\mathcal{Q}=\mathcal{P}\times L$. It is kept with a probability $Min\left(1,\; \mathcal{Q}'/\mathcal{Q}\right)$, where $\mathcal{Q}'$ is the total weight of the point being tested and $\mathcal{Q}$ is that of the source point. If the evaluated point is not kept, then a new point is generated from the last accepted point. Thus, the parameter space is scanned via a random walk by iterating this procedure.

The priors we impose are: a set of parameters has to lie within the boundaries of the parameter space given by Table~\ref{tab:par_int}, while a physical solution of the spectrum calculator and a neutralino LSP are required. Regarding likelihoods, these are displayed in Table I of~\cite{Vasquez:2010ru}. We include limits on B physics observables, on the anomalous magnetic moment of the muon $(g-2)_{\mu}$, on the Higgs and sparticles masses obtained from LEP and the corrections to the $\rho$ parameter. In the case of the limits on the Higgs mass we used the SUSY-HIT~\cite{Djouadi:2006bz} and the HiggsBounds packages~\cite{Bechtle:2008jh,Bechtle:2011sb} as in~\cite{Vasquez:2011yq} .
The HiggsBounds version used (3.1.3) include LEP and Tevatron results as well as first LHC results, while more recent results from CMS presented in ~\cite{Chatrchyan:2011nx,CMS_higgs_fit} were added a posteriori. 
 Notice that we take the WMAP measurement on the dark matter relic density as a strict upper limit on the LSP relic density -obtained via the usual freeze out mechanism-, however, we allow the neutralino to have a relic density as low as $10\%$ of the measured value. Indeed, the LSP could be only a fraction of the dark component, the rest corresponding to other dark particles or to a modified theory of gravity. For more details see~\cite{Vasquez:2010ru} and \cite{directionnelle}.

The scans we performed in this study were aimed to give a general determination of the different configurations with neutralino masses at the weak scale and below. However, as it was shown in~\cite{Vasquez:2010ru,Vasquez:2011yq}, it is difficult to find light ($\lesssim 30$ GeV) neutralinos with a random walk: 
the probability of falling in these regions that require fine-tuning is rather small.
Indeed, in the MSSM, the neutralino LSP has to annihilate via the exchange of either rather light Higgs bosons, scenarios that are heavily constrained by the Tevatron experiments as well as by CMS, or light sleptons, particularly of staus with masses close to the LEP lower bound of 81.9 GeV~\cite{Nakamura:2010zzi}. 

Hence we used two different techniques to trigger the chains that scanned the parameter spaces. On one hand we let part of the chains start randomly, i.e. look randomly for a starting point with $\mathcal{Q}\neq 0$. On the other hand we used the previous knowledge of fine-tuned regions explored in~\cite{Vasquez:2010ru,Vasquez:2011yq} to set fixed starting points for the rest of the chains, in order to force the random walk to yield at least a few points in such regions.

A summary of the characteristics of the runs we present is given in Table~\ref{tab:par_int}, for more information see~\cite{directionnelle}. 
\begin{table}[hbt]
\centering
\begin{tabular}{|c|c|c|c|}
\hline
\rm{Parameter} & \rm{Minimum} & \rm{Maximum} & \rm{Tolerance} \\
\hline
$M_1$ & 1 & 1000 & 3 \\
$M_2$ & 100 & 2000 & 30 \\
$M_3$ & 500 & 6500 & 10 \\
$\mu$ & 0.5 & 1000 & 0.1 \\
$\tan\beta$ & 1 & 75 & 0.01 \\
$M_A$ & 1 & 2000 & 4 \\
$A_t$ & -3000 & 3000 & 100 \\
$M_{\tilde{l}_R}$ & 70& 2000 & 15 \\
$M_{\tilde{l}_L}$ & 70 & 2000 & 15 \\
$M_{\tilde{q}_{1,\,2}}$ & 300 & 2000 & 14 \\
$M_{\tilde{q}_3}$ & 300 & 2000 & 14 \\
\hline
\end{tabular}
\caption{Intervals of free parameters used for the MSSM (GeV units).}
\label{tab:par_int}
\end{table}

\subsection{Astrophysical constraints}

Next we briefly describe additional constraints imposed only after exploring the parameter space with the MCMC. 
These constraints arise from two types of dark matter observables. 
First we consider the limit from the spin independent cross section for neutralino scattering on nucleons obtained by XENON100~\cite{Aprile:2011hi}. In the MSSM, the  cross section for neutralino scattering is computed with \micro~ and depends on the quark content of the nucleon as well as on the local dark matter density. To set the quark coefficients we use $\sigma_{\pi N}=45$~MeV and $\sigma_0=40$~MeV. These values lead to rather conservative astroparticle physics bounds although recent lattice QCD results~\cite{Giedt:2009mr} indicate that the s-quark content could be smaller than previously thought, leading to a further 20\% suppression of the spin independent cross sections. 
Since we allow the neutralino to be only a fraction of the dark matter component, we rescale the nominal value for the DM density, 
$\rho=0.3$,  by the fraction of the neutralino relic density to the measured relic density. Thus we have the same rescaling in all astrophysical systems. 
Second we consider the limit on the photon flux resulting from neutralino pair annihilation in dwarf spheroidal galaxies extracted from Fermi-LAT data~\cite{Abdo:2010ex}. The procedure followed was presented in ~\cite{AlbornozVasquez:2011js} and assumes a NFW profile~\cite{Navarro:1996gj}.
The predictions for both the photon flux and the direct detection spin independent rate are presented and discussed in ~\cite{Vasquez:2011yq} and ~\cite{Vasquez:2011bh,directionnelle}. Basically, the limit from  dSph's constrain the light ($<20$GeV) neutralino case, whether or not it is associated with a light pseudoscalar, while direct detection limits constrain the light neutralino/light pseudoscalar region 
as well as regions where the LSP has a large higgsino fraction. These results will be presented and discussed in a forthcoming publication~\cite{directionnelle}.

\subsection{Collider constraints}

\begin{figure}[!ht]
\includegraphics[width=8cm,height=6.5cm]{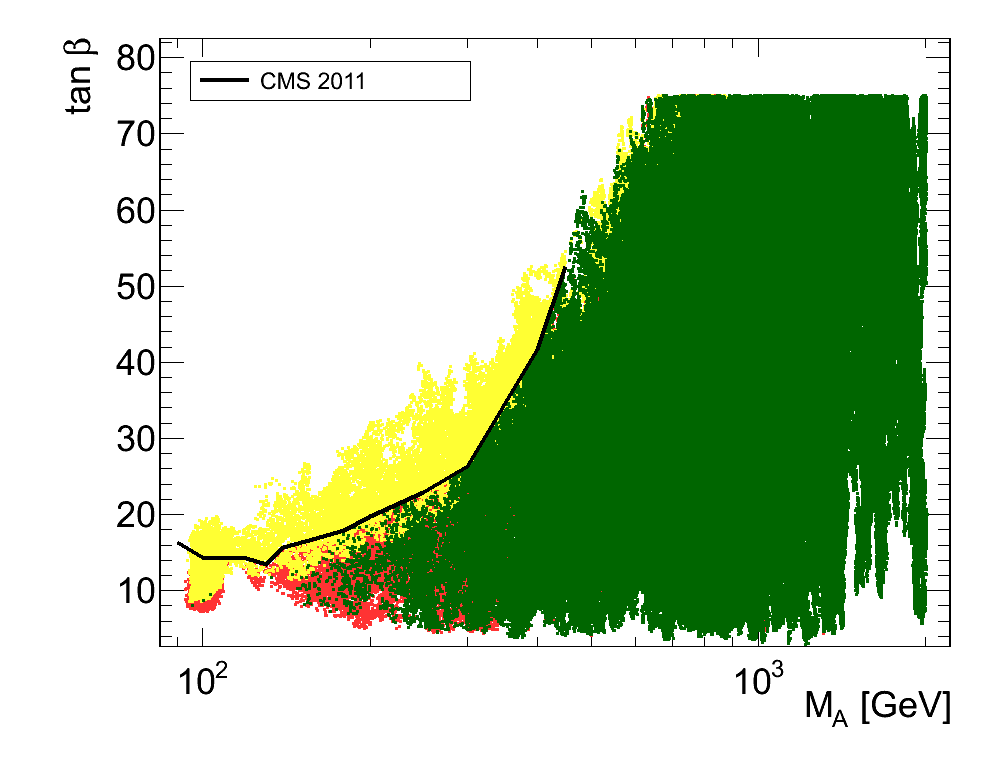} 
\includegraphics[width=8cm,height=6.5cm]{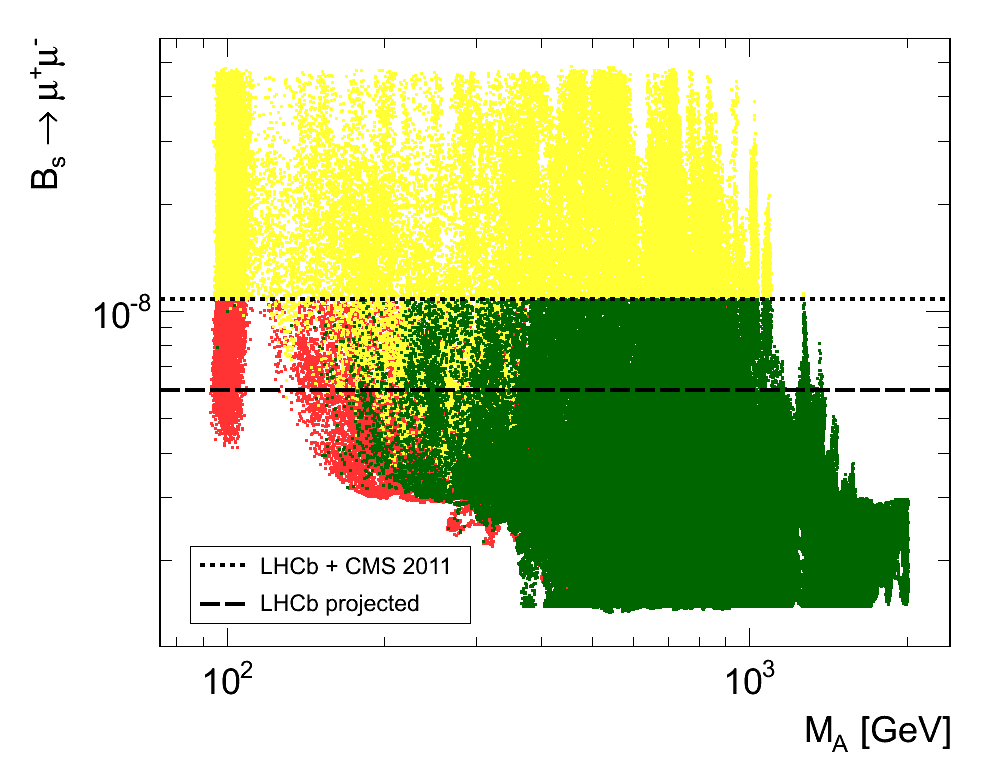} 
 \caption{Left: Allowed points in the  $\tan\beta$ vs. $M_A$ plane in green (dark grey).
 The exclusion limit from CMS~\cite{CMS_higgs20} (full line) is also displayed. 
 Right: Allowed points in the  $\bsmu$ vs. $M_A$ plane. The exclusion limit from CMS and LHCb~\cite{CMS_LHCb_bsmu} (dotted line) is also displayed. 
 In yellow, points excluded by collider constraints (Higgs and $\bsmu$), in red those excluded by astrophysical constraints  (XENON100 and Fermi-LAT). }
 \label{fig:tbmA}
\end{figure}

Since performing the parameter space exploration, new limits from the LHC were announced. The important limits on the Higgs sector are the ones for $h\ra \tau\tau$ which constrain the low $M_A$- large $\tan\beta$ region of parameter space. Furthermore the new upper limit on $\bsmu<1.08\times 10^{-8}$ ~\cite{CMS_LHCb_bsmu} also constrains the low $M_A$ region.
We have imposed both these constraints a posteriori. 
The impact of the additional constraints from LHC is displayed in Fig.~\ref{fig:tbmA} in the $\tan\beta-M_A$ plane and in the
$\bsmu-M_A$ plane.
The points excluded by either the Higgs search or $\bsmu$ are coloured in yellow (light grey). One can see that $\bsmu$ 
excludes points at lower values of $\tan\beta$ than the Higgs searches (yellow points that are below the Higgs exclusion line in Fig.~\ref{fig:tbmA} (left panel) while some points that satisfy the $\bsmu$ limit are constrained by Higgs searches (Fig.~\ref{fig:tbmA}, right panel).
 In this plot and all the following we will use the same color code: the points excluded by collider constraints are yellow, those excluded by 
either XENON100 or Fermi-LAT are red and allowed points are green (dark grey). 
Note that the scan extends to unnaturally large values of $\tan\beta$, however the special features in the observables we will discuss do not require a very large value of $\tan\beta$.

\section{Higgs observables}
\label{sec:hcoupling}

The mass of the light Higgs in the MSSM scenarios that we consider is always below 130GeV. Thus the most important detection channel is in the two-photon channel. 
In this mass range, the light Higgs is predominantly produced in gluon fusion. We define the ratio of the production times branching ratio of the MSSM to the SM,
\begin{equation}
R_{gg\gamma\gamma}= \frac{\sigma(gg\rightarrow h)_{MSSM} BR(h\rightarrow \gamma\gamma)_{MSSM}}{\sigma(gg\rightarrow h)_{SM} BR(h\rightarrow \gamma\gamma)_{SM}} 
\end{equation}
Note that $\sigma(gg\ra h)$ is taken to be proportional to $\Gamma(h\ra gg)$ even though QCD corrections are different for the two processes. We assume that the effect of QCD corrections cancels out when taking the ratio of the MSSM to the SM value. Both the MSSM and SM partial widths for $h\ra gg,\gamma\gamma$ are computed with HDECAY~\cite{Djouadi:2006bz}.

Another search channel that has been used to set powerful limits on the MSSM Higgs sector is the $\tau\tau$ channel.
Here the production is either through gluon fusion, from b-quarks, or $WW$ fusion. Thus we define three quantities for the ratio of the production times branching ratio of the MSSM to the SM
\begin{equation}
R_{gg\tau\tau}= \frac{\sigma(gg\rightarrow h)_{MSSM} BR(h\rightarrow \tau\bar\tau)_{MSSM}}{\sigma(gg\rightarrow h)_{SM} BR(h\rightarrow \tau\bar\tau)_{SM}} \;\;\; R_{bb\tau\bar\tau}= \frac{\sigma(bb\rightarrow h)_{MSSM} BR(h\rightarrow \tau\bar\tau)_{MSSM}}{\sigma(bb\rightarrow h)_{SM} BR(h\rightarrow \tau\bar\tau)_{SM}}\nonumber
\end{equation}

\begin{equation}
 R_{WW\tau\tau}= \frac{\sigma(WW\rightarrow h)_{MSSM} BR(h\rightarrow \tau\bar\tau)_{MSSM}}{\sigma(WW\rightarrow h)_{SM} BR(h\rightarrow \tau\bar\tau)_{SM}}
\end{equation}
Note that the latter is relevant for either WW fusion or associated W production.

Another search channel is the production of a Higgs in association with a gauge boson 
with the Higgs decaying into the $b\bar{b}$ final state. Similarly we define the ratio of the production times branching ratio of the MSSM to the SM
\begin{equation}
 R_{WWbb}= \frac{\sigma(WW\rightarrow h)_{MSSM} BR(h\rightarrow b\bar{b})_{MSSM}}{\sigma(WW\rightarrow h)_{SM} BR(h\rightarrow b\bar{b})_{SM}}
\end{equation}

Before showing our predictions for the observables just mentioned, we summarize the expectations for the Higgs couplings to SM particles in the MSSM. We define $R_{XXh}$ as the ratio of the Higgs coupling to a pair of SM particles ($X$) in the MSSM to the same coupling in the SM.
The $ggh$ coupling 
 is dominated by the top quark but can also receive a large contribution from the third generation squarks, in particular from the stop sector. The coupling $h\ra \gamma\gamma$ is dominated by the contribution from the W bosons with a contribution from top quarks almost an order of magnitude smaller. The two contributions have opposite signs. Supersymmetric contributions arise from the chargino and stop sector. Thus both the production and the decay can be significantly different from the SM. There are three mechanisms that can induce large corrections to some of the Higgs couplings and branching ratios: a light pseudoscalar, light supersymmetric particles in the loop and light LSP's. 
 
When  $M_A$ is light, that is in the non-decoupling case, 
 the tree-level couplings of the Higgs to SM particles can show large deviations as compared to the SM case. For example, $R_{WWh}=sin(\alpha-\beta)$ and $R_{tth}=\cos\alpha/\sin\beta$ where $\alpha$ is the Higgs mixing angle, are smaller than one only for low values of $M_A$ while   
 $R_{bbh}=R_{\tau\tau h}=\sin\alpha/\cos\beta$  are enhanced especially at large values of $\tan\beta$ as shown in ~\cite{Belanger:1999pv,Djouadi:2005gj}. Furthermore corrections to  the $hbb$ vertex arise from higher order effects, in particular the $\Delta M_b$ correction can lead to a much enhanced $hbb$ coupling at large values of $\tan\beta$~\cite{Djouadi:2005gj}. Because the $bb$ mode is the dominant decay channel of the light Higgs, in this case the total width of the Higgs becomes much larger in the MSSM. This means that the branching fraction in other modes, such as $BR(h\ra \gamma\gamma)$ can be strongly suppressed even when the $h\gamma\gamma$ coupling is itself SM-like. It is also possible that $\Delta M_b$ corrections lead to a suppressed $hbb$ coupling in which case the branching fractions into all other modes could be enhanced. However this requires large values of $\tan\beta$, low value of $m_A$ as well as large values of $\mu$~\cite{Djouadi:2005gj}, we did not find any such configurations in our MCMC analysis.

The loop-induced Higgs couplings can also deviate from their SM value because of the contributions of the supersymmetric particles in the loop.
The largest effect is from the stop sector~\cite{Djouadi:1998az}. The light stop interferes constructively with top when there is no mixing in the stop sector while interference is destructive in the large mixing case. In the first case $hgg$ will increase while in the latter it will decrease and a contrario the $h\gamma\gamma$ couplings. Since we have imposed an universal mass for the third generation, the mixing in the stop sector is large unless $A_t-\mu/\tan\beta\approx 0$ - so we expect a decrease in the Higgs production via gluon fusion, that cannot be compensated by the more modest increase of the two-photon width. Note that the largest corrections are expected for light stops, while in our scans the stops are above 400 GeV and typically above the TeV scale.
The charginos can also affect the $h\gamma\gamma$ coupling, the contribution of charginos was shown to be at most
 of the order of 15\% ~\cite{Belanger:2000tg}.

We will not consider the $WW$ decay mode of the Higgs, although the decay into virtual W's contribute to the total significance for the SM Higgs exclusion in the upper range of the MSSM light Higgs mass, it is much less important than the $\gamma\gamma$ mode. Furthermore, as discussed above we expect the $WWh$ coupling to be either SM-like or suppressed at low values of $M_A$, thus even lessening the importance of this channel.

\section{Results}

After performing the MCMC analysis and imposing the constraints described in the previous section, we examine the predictions for the light Higgs  of the model.

\subsection{The invisible Higgs }

\begin{figure}[!ht]
\includegraphics[width=8cm,height=6.5cm]{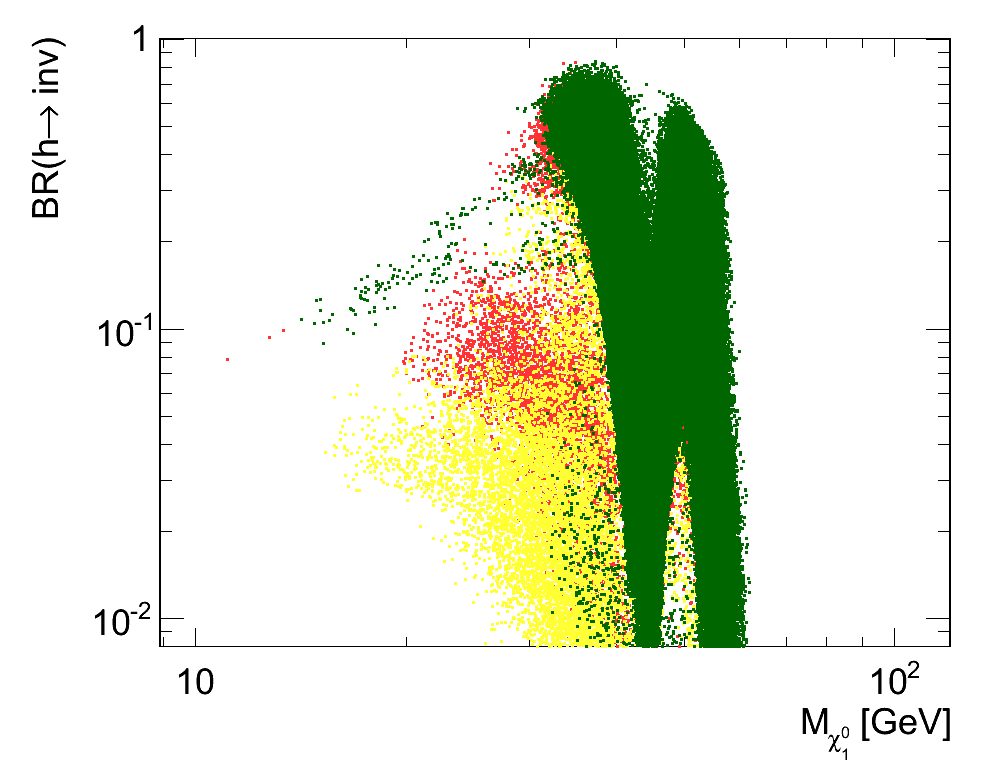} 
\includegraphics[width=8cm,height=6.5cm]{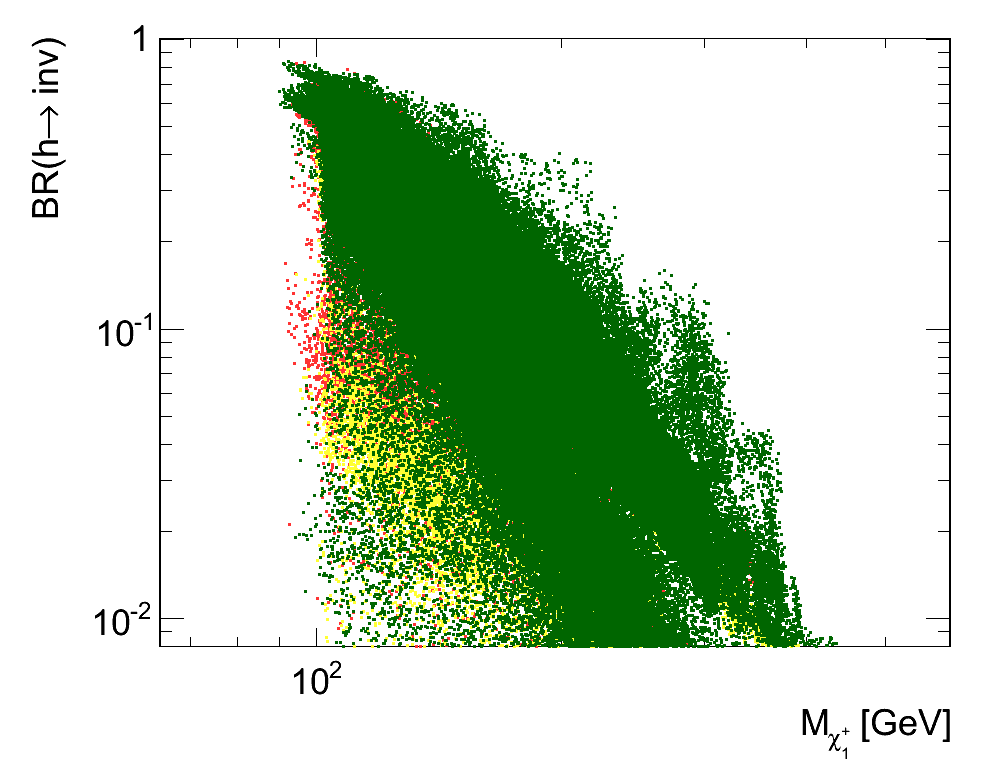} 
 \caption{ $\hinv$ as a function of the LSP mass (left) and $M_{\tilde\chi+}$ (right), same color code as Fig.~\ref{fig:tbmA}.}
 \label{fig:inv}
\end{figure}

A large branching fraction of the Higgs into invisible particles will affect all Higgs observables in the MSSM. 
Apart form the obvious kinematic condition on the LSP mass, $\mneut< M_h/2$, a large partial width into invisible requires a significant coupling of the LSP to the Higgs. This means that the light neutralino which is predominantly bino must have also a higgsino component.
Thus $\mu$ cannot be large. Since $\mu>100$~GeV because of the LEP constraints on charginos, 
this condition implies that for very light neutralinos, say below 20~GeV, the invisible width never exceeds 20-30\%. Indeed such a light neutralino cannot have a large higgsino component since $\mu\gg M_1$, see Fig.~\ref{fig:inv}, left panel.
For heavier neutralino LSP's the invisible width can reach 80\%. Note that when $\mneut\approx M_Z/2$ there is a dip in the invisible width -- it does not exceed 50\%. This is because the coupling of the LSP to the Z must be somewhat suppressed for the annihilation of the LSP via Z-exchange not to be enhanced by the resonance effect, thus avoiding a too small value for the relic density. The $\neuto\neuto Z$ coupling also depends on the higgsino component of the LSP hence $\hinv$ is suppressed at the same time. Note that a light LSP will also contribute to the invisible width of the Z. However, this constraint is taken into account in the numerical analysis.

There is a strong correlation between the mass of the lightest chargino and the invisible width. Indeed the small value for $\mu$ needed for the invisible Higgs drives the mass of the lightest chargino. We find that when the invisible width of the Higgs is larger than 20\%, then the chargino is lighter than 200GeV, see
Fig.~\ref{fig:inv}, right panel. Furthermore since these points correspond in general to $\mu<M_2$, the chargino will be largely higgsino and the second and third neutralino will also have a mass of the same order.

\begin{figure}[!ht]
\begin{center}
\includegraphics[width=8cm,height=6.5cm]{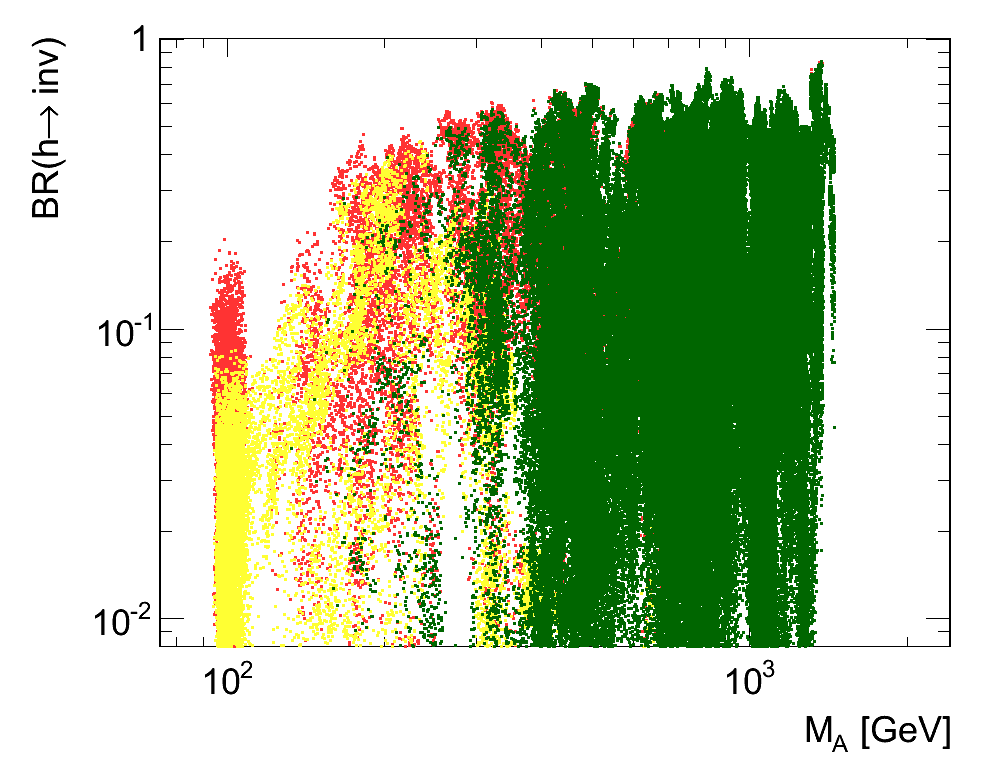} 
\end{center}
 \caption{ $\hinv$ as a function of $M_A$. Same color code as Fig.~\ref{fig:tbmA}.}
 \label{fig:inv_ma}
\end{figure}

In Fig.~\ref{fig:inv_ma}, we display $\hinv$ as a function of the pseudoscalar mass. 
The invisible branching ratio is never very large for light pseudoscalar. This is because, as mentioned in section~\ref{sec:hcoupling}, there is a large increase in the $h\ra \bar{b}b$ partial width hence also in the total width. 
Although there is no direct correlation between the invisible width and $M_A$ - in the decoupling limit any value of the invisible width can be found - this figure is included to facilitate a comparison with other Higgs observables which strongly depend on $M_A$. Note that
a large invisible width can be found even for $M_A>1$~TeV. 

The Higgs can also decay into sneutrinos. These further decay into neutralinos thus contributing to the invisible width of the Higgs.
This channel is usually not accessible kinematically since to have $M_{\tilde\nu}<M_h/2$ requires that the LH soft mass for sleptons be below roughly 100 GeV. The charged sleptons are then strongly constrained by LEP. 
Only a few points where the charged sleptons are just above the LEP limit are allowed and have a sneutrino light enough for a dominant decay of the Higgs is into sneutrinos. In the numerical analysis we have included both the neutralino and sneutrino mode in the invisible width though the LSP is always a neutralino.

\subsection{$gg\ra h\ra \gamma\gamma$ }
\begin{figure}[!ht] 
\includegraphics[width=8cm,height=6.5cm]{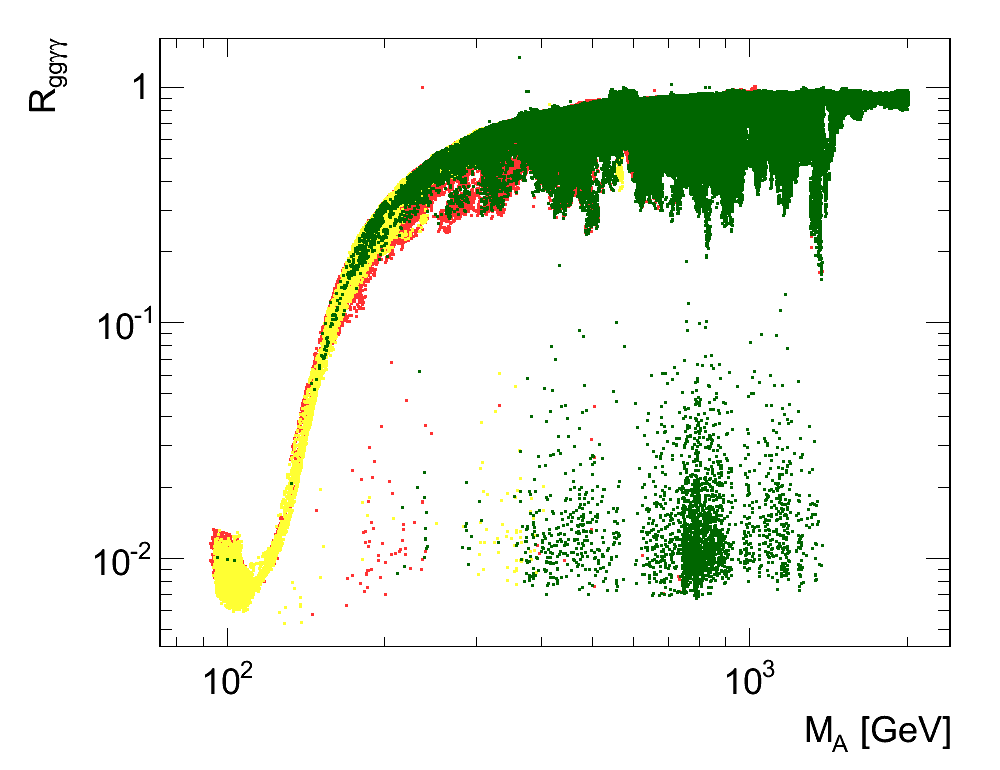} 
\includegraphics[width=8cm,height=6.5cm]{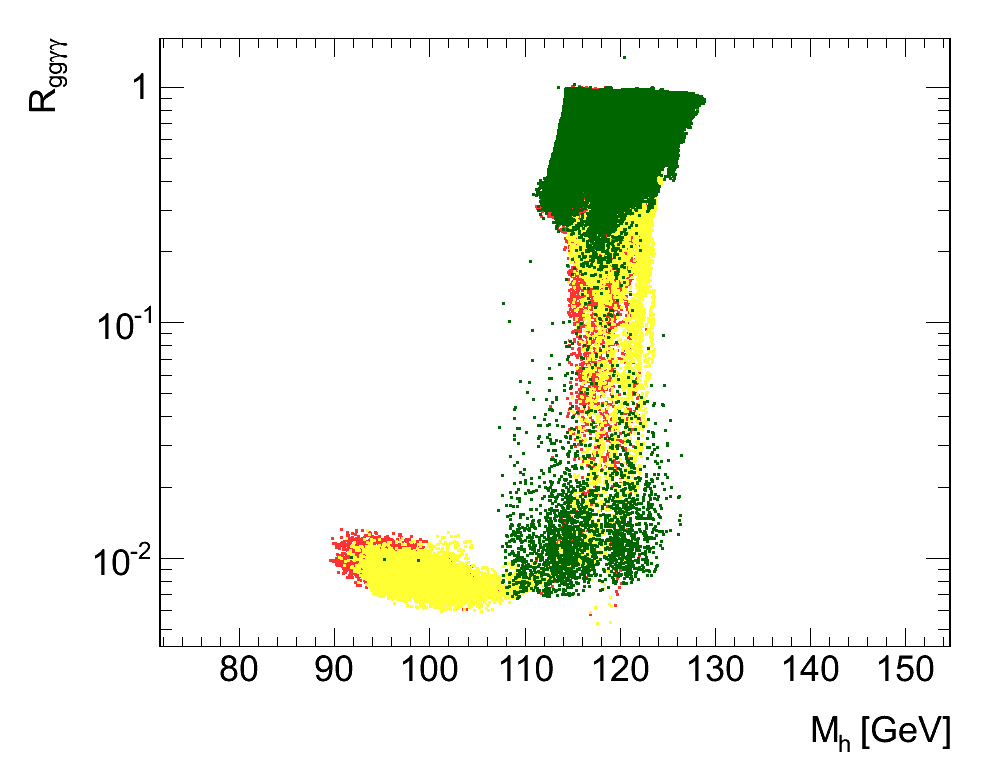} 
 \caption{ $\gggg$ as a function of the pseudoscalar mass (left) and light  scalar mass (right) for allowed points. Same color code as Fig.~\ref{fig:tbmA} }
 \label{fig:gggg}
\end{figure}

The predictions for $\gggg$ as a function of the pseudoscalar mass 
are displayed in Fig.~\ref{fig:gggg}, left panel. The MSSM Higgs signal is expected to be at most as large as that of the SM Higgs. 
The observable
$\gggg$ drops dramatically when 
$M_A$ is below about 200 GeV, as expected from the previous discussion. 
The upper limit on $\gggg$ is only 0.3(0.6) when $M_A=200(300)$~GeV.
 Although many of the points with a light pseudoscalar are already constrained by either Fermi-LAT, XENON100 (points in red), or latest Higgs searches or the new upper limit on $\bsmu$ (points in yellow ), some points with $\gggg\approx 0.01$ and $M_A\approx 100$~GeV are still allowed. 
 These points lie in the small window left by LEP constraints when all Higgses are roughly of the same mass. 
 Furthermore there are still allowed points where the Higgs signal is only 10\% of its SM value. Since 
 these are associated with a light $M_A$, they can be probed further in searches for the neutral and charged component of the heavy Higgs doublet in CMS and ATLAS and/or a more precise measurement of $\bsmu$. Note that one needs to explore values of $\tan\beta<10$ to be able to probe these points. 

When $M_A>400$~GeV and one is in the decoupling limit the suppression in $\gggg$ is usually more modest. 
Nevertheless it is still possible to have $\gggg$ as low as 0.2 even when $M_A$ is at the TeV scale. 
This occurs when the LSP is lighter than $M_h/2$ and the invisible width of the Higgs is large.
Note that there are a few points at large values of $M_A$ where $R\approx 0.01$ these all correspond to points where the LH soft mass for sleptons is below 100 GeV, the charged sleptons are just above the LEP exclusion and the sneutrino is light enough that the dominant decay of the Higgs is into sneutrinos.
Other suppression of $\gggg$ arises from the stop sector. The effect is usually below 30\%, although for stop masses below 500 GeV we found points where $\gggg$ can drop as low as 0.4.

The predictions for $\gggg$ as a function of the light Higgs mass show that $\gggg$ has little dependence on the scalar mass provided it lies in the $110-130$~GeV range. However it is important to note that for a Higgs in the narrow mass range $110-114$~GeV, we always find $\gggg<0.8$, because these values of $M_h$ are found only in the non-decoupling limit.
Finally as explained above, the few allowed points around $M_h=100$~GeV feature a large drop in the Higgs two-photon signal.

\subsection{$h\ra f\bar{f}$}

\begin{figure}[!ht]
\includegraphics[width=8cm,height=6.5cm]{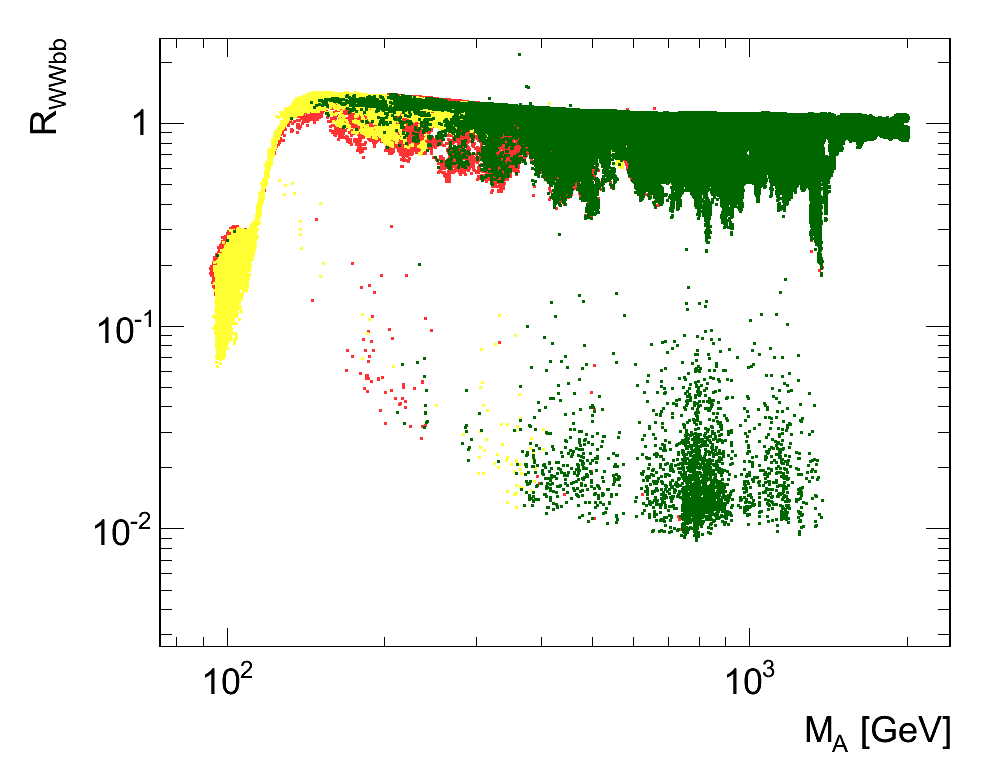} 
\includegraphics[width=8cm,height=6.5cm]{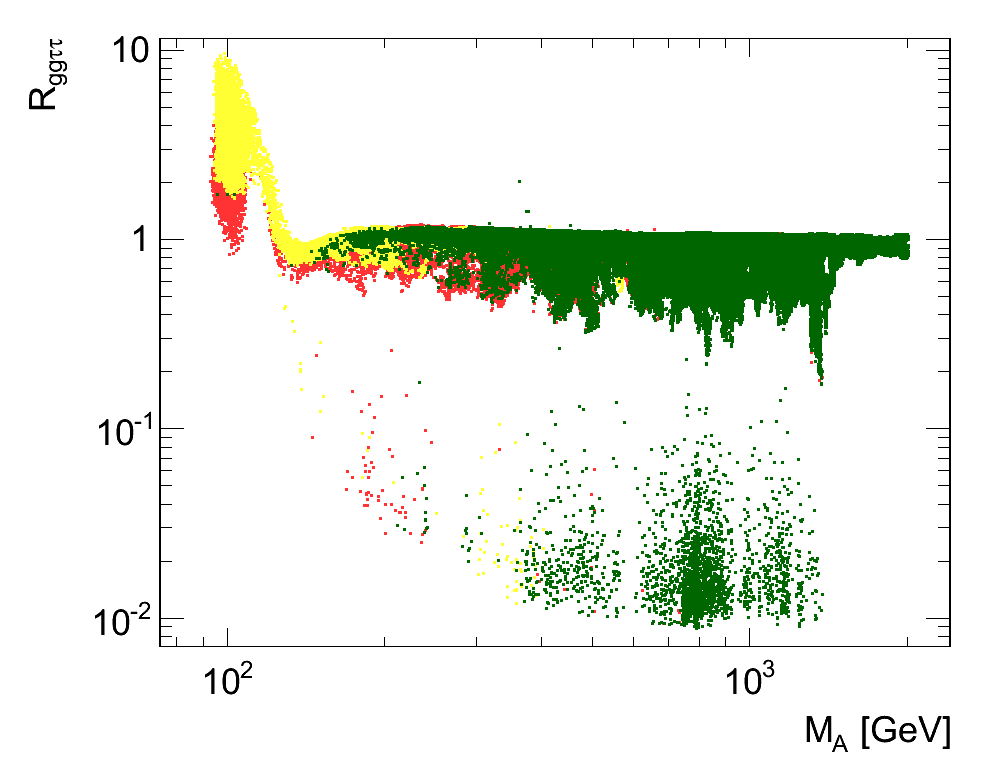} 
 \caption{ $R_{WWbb}$ (left) and $\ggtau$ (right) as a function of the pseudoscalar mass for allowed points. Same color code as Fig.~\ref{fig:tbmA} }
 \label{fig:hll}
\end{figure}

Higgs searches are performed for two different final states, $bb$ and $\tau\tau$.
First consider associated $Wh$ production with the decay $h\ra b\bar{b}$. In the decoupling limit, both the coupling $hWW$ and $hbb$ are near their SM value, thus we find that $R_{WWbb} \approx 1$ when $M_A>400$~GeV except for the points with a large invisible width. 
Since $b\bar{b}$ is the dominant decay mode of the Higgs in both the MSSM and the SM, there can only be a modest enhancement of the branching fraction even when the coupling is much enhanced. 
The largest enhancement of $BR(h\ra b\bar{b})$ is found when $M_A<400$~GeV. As mentioned in section~\ref{sec:hcoupling}, this leads to 
a value $R_{WWbb}$ that can reach 1.2 when the coupling $hWW$ is close to the SM value,see Fig.~\ref{fig:hll}. However, when $M_A$ is lighter than 200~GeV, the $WWh$ coupling is strongly suppressed and $R_{WWbb}$ can drop to values as low as 0.1.

Next we consider the signal in $h\ra \tau\bar\tau$ treating separately three production processes. 
The dominant production process is gluon fusion. 
We find that $R_{gg\tau\tau}$ can reach almost 10, but only for values of $M_A$ that are constrained from heavy Higgs, $\bsmu$ searches and/or from astrophysical constraints, see Fig.~\ref{fig:hll} (right panel). The few allowed points around $M_A\approx 100$~GeV have 
$R_{gg\tau\tau}\approx 2$  due an  increase of the $ggh$ coupling resulting from a much enhanced
contribution of  b quarks in the loop. 
Otherwise
$R_{gg\tau\tau}$ does not deviate much from the SM (0.8-1.1) unless there is a large invisible width. 

\begin{figure}[!ht]
\includegraphics[width=8cm,height=6.5cm]{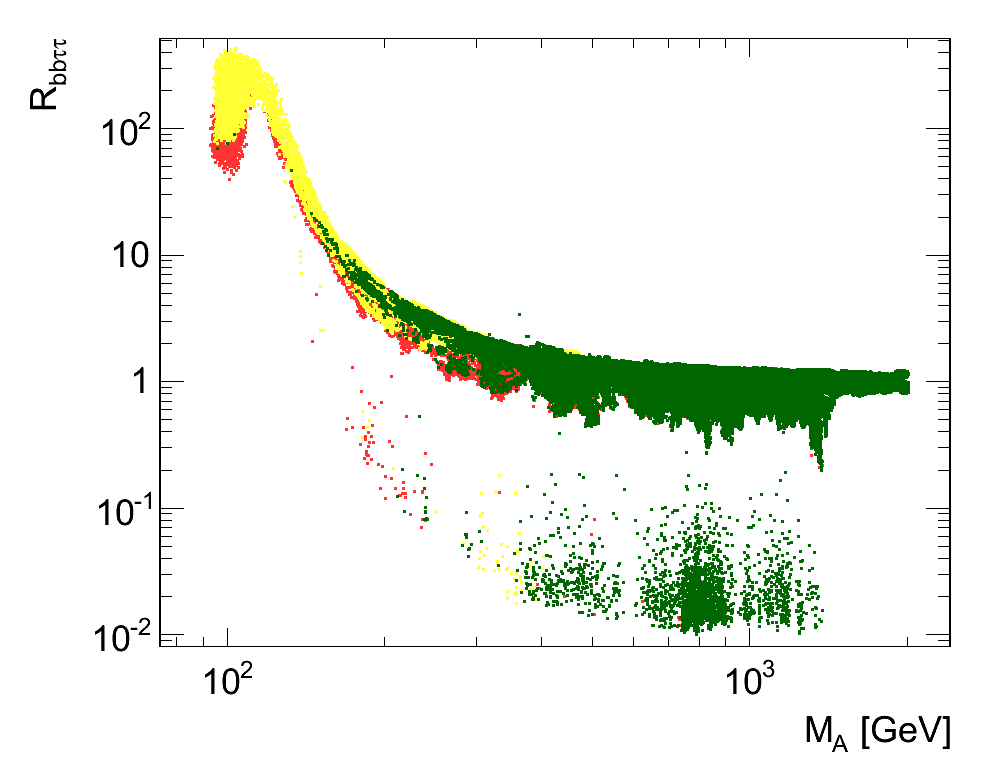} 
\includegraphics[width=8cm,height=6.5cm]{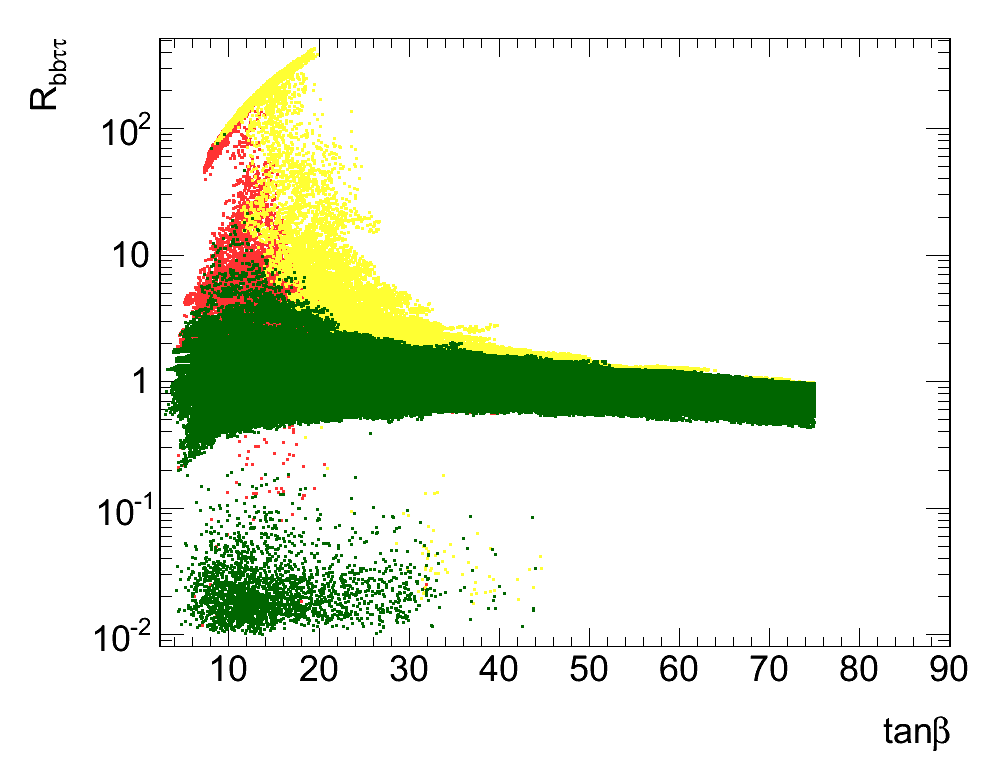} 
 \caption{ $\bbtau$ as a function of the pseudoscalar mass (left) or $\tan\beta$ (right) mass for allowed points. Same color code as Fig.~\ref{fig:tbmA} }
 \label{fig:bbhll}
\end{figure}

The second process with Higgs production  via W fusion and decaying into 
$\tau \tau$ final states is driven by $R_{WW\tau\tau}$. This quantity behaves very much like $R_{WWbb}$ discussed above. Indeed, at tree-level  the branching fraction into tau pairs is proportional to the $b\bar{b}$ branching. Supersymmetric radiative corrections can induce shifts in   the $hbb$ coupling, however we have checked that for the set of allowed points, the $\tau\tau$ and $bb$ branching ratios are well correlated. 
The third process is Higgs production from $b$-quarks. It is for this channel that we find the most dramatic effect as compared to the SM expectations. Indeed  for small values of $M_A$ 
 the $hbb$ coupling can be increased, section~\ref{sec:hcoupling}, thus leading to up to $R_{bb\tau\tau}\approx 100$, see Fig.~\ref{fig:bbhll}. 
Such large enhancements are possible even for intermediate values of $\tan\beta$. In fact for very large values of $\tan\beta$, $R_{bb\tau\tau}$ is at most equal to the SM value, see right panel of Fig.~\ref{fig:bbhll}. This is because $M_A$ is strongly constrained from LHC searches when $\tan\beta$ is large.
Altogether the $bb\tau\tau$ channel is mostly enhanced as compared to the SM, apart from the points with a significant invisible width,
although the deviation from the SM prediction  is quite modest when $M_A>400$~GeV, see Fig.~\ref{fig:bbhll}. 
Note however that the SM production rate of the Higgs from b quarks is very small as compared with W and gluon fusion.

\subsection{Correlation between Higgs search channels}

\begin{figure}[!ht]
\includegraphics[width=8cm,height=6.5cm]{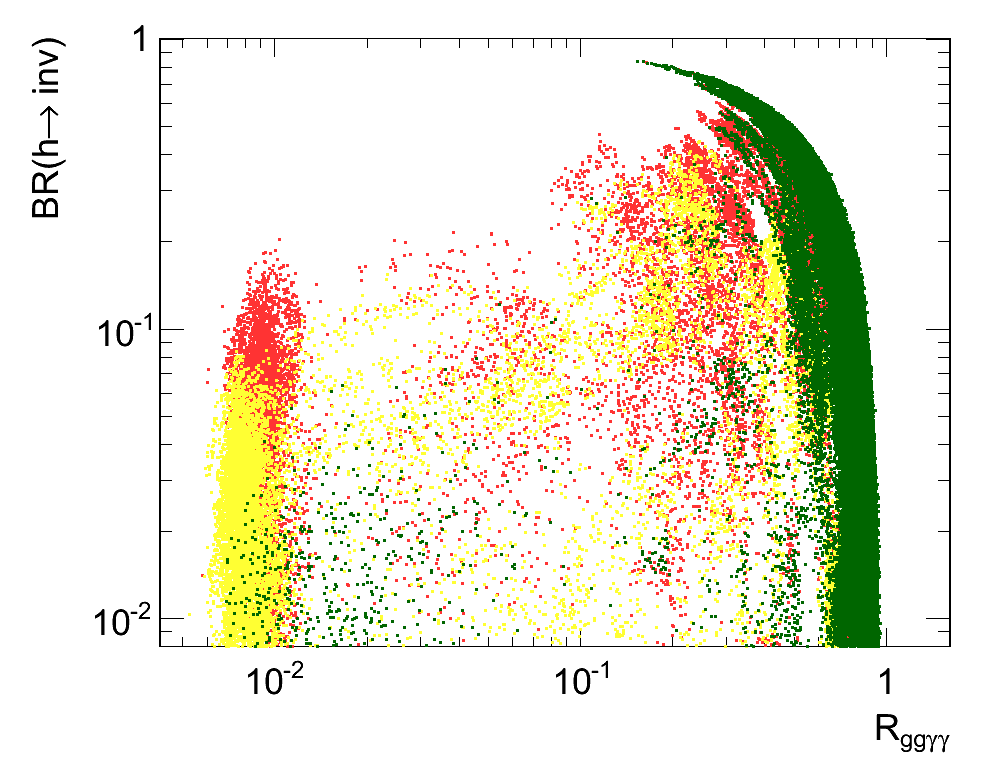} 
\includegraphics[width=8cm,height=6.5cm]{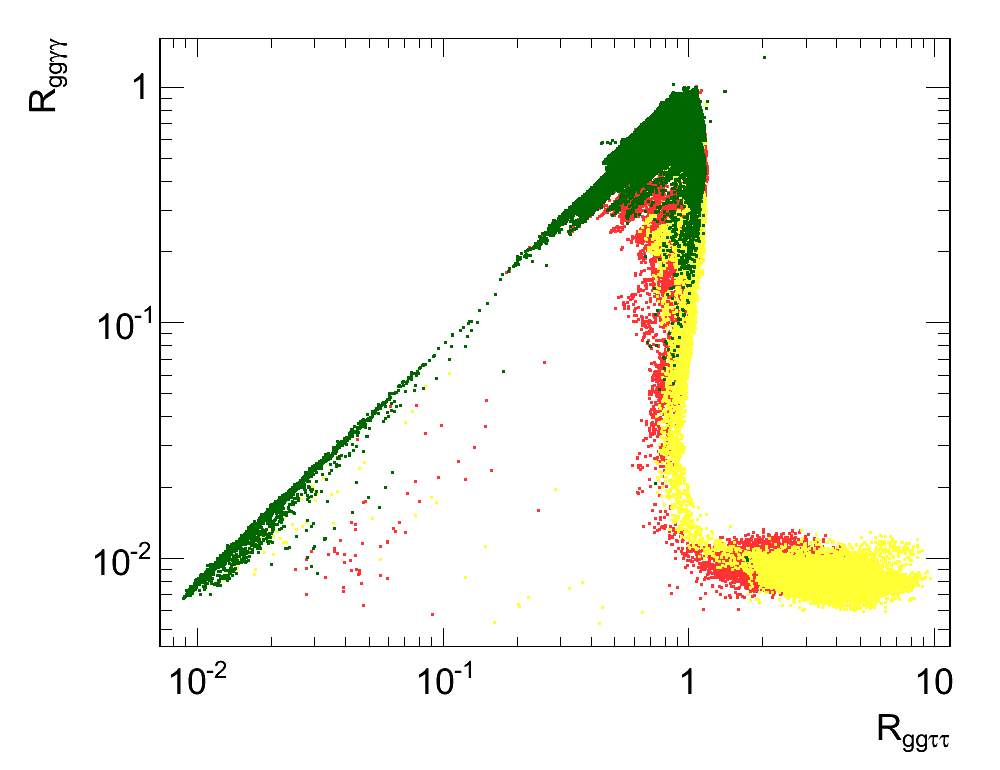} 
 \caption{Allowed points in the plane $\hinv$ vs $\gggg$ (left) and $\gggg$ vs $R_{bb\tau\tau}$, same color code as Fig.~\ref{fig:tbmA} }
 \label{fig:gg_inv}
\end{figure}

The correlation between the large invisible width and the suppressed $\gggg$ is displayed in Fig.~\ref{fig:gg_inv}, left panel. 
When $\hinv$ is at the percent level, one sees a large allowed band where $\gggg>0.7$. As $\hinv$ increases, the value of $\gggg$ drops to less than 20\%.
To the left of this band there are a good number of points which have $\gggg$ between 0.3-0.7 even when the invisible width is small. 
These correspond to values of $M_A < 300$~GeV where the $BR(h\ra \gamma\gamma)$ is suppressed because of a large enhancement of the total width of the Higgs. 
Finally there are a number of scattered points with very suppressed $\gggg$ corresponding to the small values of $M_A$. 
Despite the fact that the production process is the same, there is no direct correlation between the channels $\gggg$ and $\ggtau$ only when the Higgs decays invisibly, this is the branch where both channels are suppressed in Fig.~\ref{fig:gg_inv}. There are scenarios with a strong suppression in $\gggg$ where $\ggtau$ is near or above 1.
Again this is due to the large enhancement of the total width of the Higgs due to a large $hbb$ coupling.
We therefore also find in this case $R_{WWbb}$ to be near 1. 
Conversely there are cases where $\gggg \approx 1$ and $\ggtau$ is suppressed, see Fig.~\ref{fig:gg_inv}. Note that these points should all be probed in the most sensitive channel, $gg\ra h\ra\gamma\gamma$, since a cross section at the SM level will be probed in the near future~\cite{Higgs125,CMS_higgs20}

To analyse the impact of suppressing the signal in one channel, we have selected the points with $\gggg<0.8$. We have also imposed 
 $M_A>200$~GeV because we have already argued that the light pseudoscalar region would be probed further by the $H/A/h\ra\tau\tau$ search. We find that when $\hinv$ is at the percent level, $M_A<500$~GeV and $\ggtau,R_{WW\tau\tau},R_{WWbb}$ are all greater than 1, 
 in agreement with our argument that the reduction in $\gggg$ is largely due to the increase of $hbb$. Furthermore we also find that $\bbtau$ is enhanced.
 When $\hinv$ increases, all channels can be reduced. However it is only when the invisible width becomes large that new possibilities to probe the Higgs through the invisible mode can be used. Furthermore luminosities larger than 10$fb^{-1}$ might be required~\cite{deroeck,Eboli:2000ze,Godbole:2003it}.

\subsection{Spectrum of superparticles}

As we have seen in the previous section, there are points that fit all constraints and where one/all the Higgs discovery channels are suppressed. As we have discussed above, the suppression of some Higgs channel is often tied to the presence of not too heavy supersymmetric particles and/or pseudoscalar/charged Higgs. 
In such a case, direct searches for supersymmetric particles are important and complementary to the light Higgs searches. As an example, we consider the case where the main Higgs channel $\gggg<0.8$ and $M_A>200$~GeV, and examine the impact on the supersymmetric spectrum. The distribution of the chargino mass is strongly shifted towards lower values $\mchar<200$~GeV. As argued above this is because the drop in $\gggg$ is mostly due to the invisible width which to be large requires a non-negligible higgsino content in the LSP, hence small values of $\mu$ and $\mchar$. 
The slepton mass distribution is also shifted towards lower values. Even though the slepton mass does not in general directly enter the Higgs production and decay, selecting suppressed $\gggg$ means including all points with a large invisible width thus with a light bino LSP and a light chargino. The constraint from $(g-2)_\mu$ is then easily satisfied either with the neutralino smuon contribution which requires not too heavy smuons or from the chargino/sneutrino contribution which requires large values of $\tan\beta$ and light sneutrinos~\cite{Martin:2001st,Byrne:2002cw}. In both cases light sleptons are preferred. Note 
that a large invisible width of the Higgs into sneutrinos is associated with $M_{\tilde{l}} \approx 100$~GeV.
Finally the stop quark mass distribution spans the whole range of allowed values thus proving to be uncorrelated to this selection. 

Searches for purely electroweak particles, charginos, neutralinos and sleptons will therefore offer a powerful and complementary probe of Supersymmetry in cases where the light Higgs is hidden.
A more quantitative analysis of the potential of the LHC to probe these channels is kept for a future work.

\begin{figure}[!ht]
\includegraphics[width=8cm,height=6.5cm]{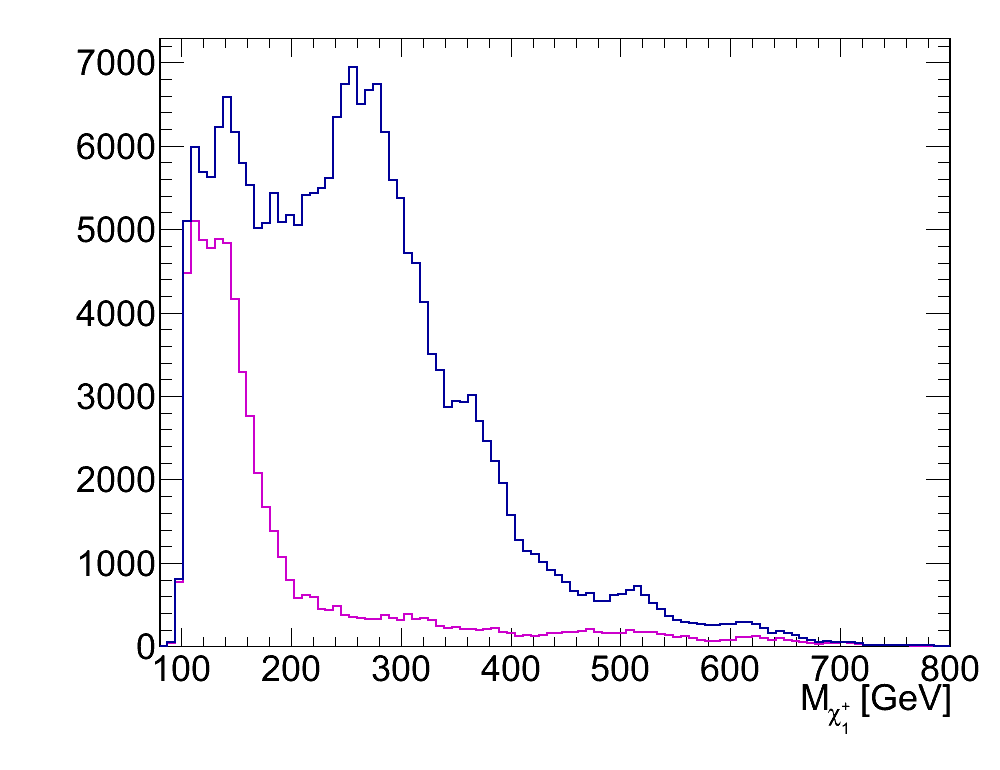} 
\includegraphics[width=8cm,height=6.5cm]{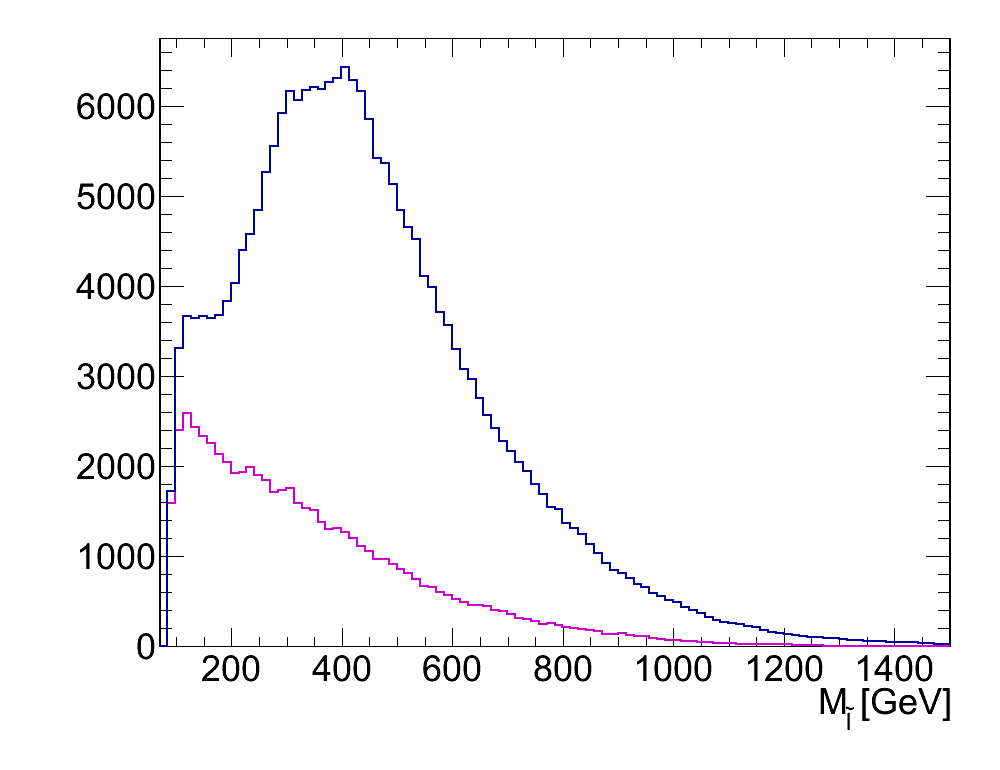} 
 \caption{ The distribution for the lightest chargino(left) and lightest slepton(right) mass for all allowed points (top,blue) and for the case where $\gggg<0.8$ and $M_A>200$~GeV (bottom,purple). }
 \label{fig:masses}
\end{figure}

\subsection{Benchmarks}

\begin{table}[h]
\centering
\begin{tabular}{|c|c|c|c|c|c|c|}
\hline
\rm{Parameter} & \rm{A} & \rm{B} & \rm{C}& \rm{D} & \rm{E} & \rm{F} \\
\hline
$M_1$ & 55 & 144.2 &211&48.0& 52.2 & 41.2\\
$M_2$ & 628 & 518  &1426&221&223& 186\\
$M_3$ & 1648 & 4762 &3745&5596&4794& 4059\\
$\mu$ & 103.2 & 114.6 &200&132.2&124.1& 707\\
$\tan\beta$ & 9.95 & 12.0 &5.0&9.77&6.19& 9.75 \\
$M_A$ & 97.8 & 265 &500&409&1353& 747 \\
$A_t$ & -1014 & -2886 &-2091&127&2112& -390 \\
$M_{\tilde{l}_L}$ &114 & 157  & 239&235&197& 77.3\\
$M_{\tilde{l}_R}$ & 632 & 765 &2023&937&1657& 1017\\
$M_{\tilde{q}_{1,\,2}}$ & 592.3 & 1946 &1471&1789&1514& 1505 \\
$M_{\tilde{q}_3}$ & 638.8 & 1403  &1099&1812&1959& 1313\\
\hline
$\mneut$ &40.1&89.6& 168&38.0&37.9& 39.3\\
$\mchar$ & 101.2   & 112.6  & 202  &111.8  & 102.5 & 189.2 \\
$M_h$ &94.2&123.5&116.7&114.3&118.1&110.9\\
$R_{hgg}$ &1.73&0.79&0.762&0.926&0.933&0.949 \\
$R_{hbb}$&82.2&2.04&1.19&1.29&1.024&1.07\\
$R_{hWW}$&0.193&1.0&1.0&1.0&1.0&1.0\\
$\hinv$&0.04&0.0& 0.0 &0.532&0.734&0.99\\
${\rm BR}(h\ra\tau\bar\tau)$/SM&1.03&1.17&1.02&0.474&0.261&0.0088\\
$\gggg$&0.008&0.476 &0.704&0.358&0.257&0.0071\\
$R_{WWbb}$&0.197&1.17&1.02&0.473&0.261&0.0088\\
$R_{gg\tau\tau}$&1.78&0.922&0.775&0.439&0.243&0.0088\\
$R_{WW\tau\tau}$&0.199&1.17&1.02&0.474&0.261&0.0088\\
$R_{bb\tau\tau}$&84.8&2.38&1.21&0.614&0.267&0.0094\\\hline
\end{tabular}
\caption{Benchmarks in the MSSM with some suppressed Higgs signal, for Benchmark F, $M_{\tilde\nu}=44.4~{\rm GeV}$. These are selected from the set of allowed points. }
\label{tab:bench}
\end{table}

To illustrate the different possible scenarios for the main Higgs search channels that we have discussed in the previous section, we list a few benchmarks in Table~\ref{tab:bench} . The first three benchmarks have a small invisible width, yet $\gggg<1$ while the last three have a large invisible width. 
Benchmark A is an example with $M_A<100$~GeV where the $h\ra b\bar{b}$ partial width and hence the total width is strongly enhanced, leading to suppressed branching ratios for $h\ra\gamma\gamma$. 
This scenario features squarks below 700 GeV which are easily probed at the LHC. 
Benchmark B  feature a heavier pseudoscalar, yet the $hbb$ coupling is enhanced, leading again to a suppressed $BR(h\ra\gamma\gamma)$, 
furthermore supersymmetric particles reduce the $hgg$ vertex. 
Benchmark C is an example where the main effect comes from the stop loop in the $hgg$ vertex.
Benchmark D and E are two examples where the Higgs decays invisibly, for two different pseudoscalar masses. Both cases feature a light chargino which is dominantly higgsino. 
Benchmark F is an example where the sneutrino is light, $M_{\tilde\nu}=44.4$~GeV, the invisible Higgs is near 100\% and all channels are suppressed. This scenario is best probed by slepton searches. 
The benchmarks illustrate explicitly the fact that the modification of the SM rate is not the same in each channel, even for a common final state. As the sensitivity of the Higgs searches improve, it would be useful to provide the results for each production times decay channel separately.

\section{Discussion}

We have displayed the predictions for several search channels. We briefly compare our results with the expectations from the LHC.
The combined ATLAS and CMS limits for the $h\gamma\gamma$ channel exclude $\sigma/\sigma_{SM}<3.1$ in the mass range $100-121$~GeV and $\sigma/\sigma_{SM}<1.8$ in the $121-131$~GeV range. 
For the higher mass range, one expects to reach $\sigma/\sigma_{SM}\approx 1$ 
with the integrated luminosity ${\cal L}=5fb^{-1}$ collected in 2011, assuming a naive rescaling of the current limit with 
$\sqrt{{\cal L}}$. This is compatible with the forecast of the CMS and ATLAS collaborations. A further increase in luminosity to 10$fb^{-1}$ would allow to probe a significant fraction of the supersymmetric parameter space through the light Higgs search in $\gamma\gamma$, since it would cover roughly the case where $\gggg>0.7-0.8$. To probe a Higgs below 120 GeV  will require at least 10$fb^{-1}$ in the $h\gamma\gamma$ channel, and even more in the case $110~{\rm GeV}< M_h<114$~GeV  since not only the exclusion is not as stringent but also the signal is expected to be suppressed as compared to the SM. 
A luminosity of $10fb^{-1}$ might be sufficient to start excluding the Higgs in the $\tau\tau$ channel if $M_h<125$GeV, especially when there is an enhancement over the SM expectation. In fact for $M_h\approx 120$GeV both the $\gamma\gamma$ and
$\tau\tau$ channels are competitive in the scenarios where the latter is enhanced. 
The case $110~{\rm GeV}< M_h<114$~GeV as well as the main channel will require a much higher sensitivity since a
 suppression occurs for the $\tau\tau$ and $b\bar{b}$ channels. 
A complete investigation of the MSSM light Higgs calls for a dedicated search for the invisible Higgs, as a large invisible width is allowed provided the neutralino LSP (or the sneutrino) is light.

A complete exploration of the parameter space of the MSSM is a daunting task. Nevertheless, even though we restricted our analysis to the MSSM with 11 parameters we think we have covered the essential features of the Higgs signals in different channels, especially concerning the difficult scenarios where the main channel is suppressed. 
In this study we have restricted ourselves to the case of common left and right stop masses, which automatically gives a large mixing among the third generation squarks, and have imposed soft squark masses above 500 GeV. 
If the mixing in the stop sector is small, $ggh$ will in fact increase as discussed in section~\ref{sec:hcoupling} leading to a stronger signal in the most sensitive search channel. These scenarios will therefore be well covered by Higgs searches. 
We might nevertheless have missed some light stop scenarios with a large mixing which have a suppressed production rate in the $gg$ mode. These should be probed by direct searches for light stop quark at the LHC~\cite{Belanger:1999pv}.

 We have also restricted our analysis to $\Omega h^2>10\%\Omega _{WMAP}$ for a neutralino LSP as the DM candidate. This means that we have excluded scenarios where the neutralino LSP has a large higgsino or wino component and constitutes a negligible component of the dark matter. 
The most significant effect of a small value of $\mu$, thus large higgsino component,
 implies a large invisible width of the Higgs when the neutralino LSP is light. This case was  however explored thoroughly
 in our study since the light LSP has typically a large value of $\Omega h^2$.
 The value of $\mu$ can impact the loop induced couplings of the Higgs, however the effect is typically below the 20\% level, similar to what we have found in the the scenarios considered in our analysis. 
 Note that heavier LSP's with large higgsino component are also strongly constrained by direct detection searches. we have also checked that by restricting $\Omega h^2$ to be compatible with the value measured by WMAP does not affect significantly our predictions for  the light Higgs production and decays.

 In conclusion our analysis  shows that when taking into account dark matter and flavour constraints in the MSSM, the light Higgs signal in the $\gamma\gamma$ channel is expected to be at most at the level of the SM Higgs. 
 These results agree with another analysis in the case where the invisible width is small~\cite{Arbey:2011un}.
 Strong suppression of the signal occur in two different cases: low $M_A$ or large invisible width. If such strong suppression occur, one must rely on alternate channels, such as the search for the heavy pseudoscalar, the charged Higgs or supersymmetric particles. A more modest suppression is associated with the effect of light supersymmetric particles.
 Observation of a SM-like Higgs signal in the mass range compatible with MSSM predictions would therefore strongly constrain the light pseudoscalar as well as light neutralino MSSM scenarios.
 Therefore new particles searches and Higgs searches provide complementary probes of supersymmetry.

\section{Acknowledgements}

We thank Guillaume Drieu La Rochelle, Fawzi Boudjema, C\'eline Boehm and Sabine Kraml for useful discussions.
GB thanks the LPSC, Grenoble and RMG thanks the LAPTH for their hospitality. 
This work was supported in part by the GDRI-ACPP of CNRS.
The work of AP was supported by the Russian foundation for Basic Research, 
grant RFBR-10-02-01443-a. RMG  wishes to acknowledge the Department of Science and Technology of India,
for financial support  under the J.C. Bose Fellowship scheme under grant no.
SR/S2/JCB-64/2007.

\providecommand{\href}[2]{#2}\begingroup\raggedright
\endgroup

\end{document}